\documentclass[a4paper,11pt]{article}

\usepackage{jinstpub} 

\title{Real-time Inference with 2D Convolutional Neural Networks on Field Programmable Gate Arrays for High-rate Particle Imaging Detectors}

\author[a,*]{Y.~Jwa,\note{Corresponding author.}}
\author[a]{G.~Di Guglielmo,}
\author[a]{L.~Arnold,}
\author[a]{L.~Carloni,}
\author[a]{G.~Karagiorgi}

\affiliation[a]{Columbia University, New York, NY, 10027, USA}

\emailAdd{yj2429@columbia.edu}

\abstract{We present a custom implementation of a 2D Convolutional Neural Network (CNN) as a viable application for real-time data selection in high-resolution and high-rate particle imaging detectors, making use of hardware acceleration in high-end Field Programmable Gate Arrays (FPGAs). To meet FPGA resource constraints, a two-layer CNN is optimized for accuracy and latency with KerasTuner, and network \textit{quantization} is further used to minimize the computing resource utilization of the network. We use ``High Level Synthesis for Machine Learning'' (\textit{hls4ml}) tools to test CNN deployment on a Xilinx UltraScale+ FPGA, which is a proposed FPGA technology for the front-end readout system of the future Deep Underground Neutrino Experiment (DUNE) far detector. We evaluate network accuracy and estimate latency and hardware resource usage, and comment on the feasibility of applying CNNs for real-time data selection within the proposed DUNE data acquisition system.}


\begin{document}
\maketitle
\flushbottom

\section{Introduction}
\label{sec:intro}

Modern-day particle physics experiments are known to produce a vast amount of data that ultimately must be reduced by employing algorithms to preferentially select only those kinds of (usually rare) signals that can be deemed as potentially interesting for further physics study and scientific discovery. This process of data selection is typically applied across several stages of the data processing pipeline, using algorithms that increasingly make use of deep learning \cite{Radovic:2018dip,Karagiorgi:2021ngt}. However, as data rates grow, there is increased motivation to accurately and efficiently execute data selection in real time, i.e.~at a rate commensurate with data throughput and with low latency, by employing ``triggers''. These are data-driven decisions, which translate physical measures---quantities calculated based on the incoming data itself and/or other external signals---into instructions on which data to keep and which data to discard. 

More recently, driven in part by the need to increase accuracy in selecting high-dimensional and highly-detailed data from modern-day particle detectors, machine learning (ML) algorithms based on both supervised and unsupervised learning have been proposed and shown to be capable of effectively triggering on incoming physics data, proving to be a viable solution for the upcoming data challenges of future particle physics experiments (see, e.g.~\cite{Jwa:2019zlh,Diotalevi:2021vlw,Elabd:2021lgo,Aad:2021tru,Govorkova:2021utb,Mikuni:2021nwn,Deiana:2021niw}). Implementing ML algorithms into dedicated hardware for triggering, such as GPUs, or FPGAs, can potentially guarantee fast execution of the algorithm while taking advantage of the algorithm's accuracy in selecting data of interest with maximal signal efficiency and signal purity. Additionally, software toolkit development projects such as \textit{hls4ml} \cite{Fahim:2021cic} are providing suitable and user-friendly frameworks for easily employing ML algorithms into hardware for application-specific usage (see, e.g.~\cite{Loncar:2020hqp,Aarrestad:2021zos}).

Motivated by a widely used particle imaging detector technology---that of liquid argon time projection chambers (LArTPCs)---we explore the applicability of algorithms commonly used in image analysis for triggering purposes. LArTPCs work by continuously imaging a large and homogeneous 3D detector volume, wherein electrically charged particles are visible through the trails of ionization they leave along their trajectories. This type of technology is increasingly employed in searches of rare events such as interactions of dark matter particles or supernova core-collapse neutrinos with the detector medium. More so than for other particle detector technologies, LArTPC data are well-suited for image analysis given that neutrino or other rare event signals are translationally invariant within a generally sparse 2D or 3D image. Indeed, in past work we have shown that 2D convolutional neural networks (CNNs) tested on simulated raw data from a LArTPC can yield sufficient accuracy and can be implemented onto parallelized data processing pipelines using GPUs to perform data selection in a straightforward way, while meeting the physics performance and latency requirements of future LArTPC experiments \cite{Jwa:2019zlh}. 
The need to further improve the long-term operation reliability and power utilization of such data processing pipelines motivates the exploration of alternate implementations of CNN-based data selection, specifically implementations on Field Programmable Gate Arrays (FPGAs).

FPGAs are devices commonly used in front-end readout electronics systems for particle physics experiments; their on-device nature (often capable of receiving the full-rate of detector-generated data prior to any data filtering or reduction) and their reliability for long-term operation make them particularly attractive for data processing algorithm implementation, especially if only minor pre-processing is necessary in the data pipeline. In general, algorithm implementation into a front-end device is advantageous as it makes large data movement unnecessary, reduces power consumption and trigger latency, and increases reliability. In this paper, we investigate the implementation of a relatively small 2D CNN onto an FPGA that is specifically targeted for use in the front-end readout electronics of the future Deep Underground Neutrino Experiment (DUNE) \cite{DUNE:2020lwj,DUNE:2020mra,DUNE:2020txw,DUNE:2020ypp}, following preliminary exploration in \cite{Jwa:2019zlh}. Keeping in mind the 2D nature and high resolution of LArTPC raw data, we explore and evaluate techniques to reduce the computational resource usage of the CNN inference on the target FPGA, in order to meet the technical specifications of the DUNE readout system, while still satisfying the physics accuracy requirements of the experiment.

The paper is organized as follows: In Sec.~\ref{sec:dune}, we describe the DUNE Far Detector (FD) LArTPC in more detail, including its operating principle, and the technical specifications and requirements of its readout and data selection (trigger) system. In Sec.~\ref{sec:CNNDS} we explore different CNN architectures, and explore their accuracy in selecting data containing rare signal events, paying particular attention to the overall size of the network, in anticipation of minimal computational resource availability in the DUNE FD readout system. Subsection~\ref{subsec:prep} describes how simulated raw data from the DUNE FD are prepared as input to the CNN; subsection~\ref{subsec:CNN_DS} describes some CNN architectures and the classification accuracy performance on simulated input images; in subsection~\ref{subsec:auto}, we further optimize the network architecture and hyperparameters in an automated way, using the KerasTuner package \cite{omalley2019kerastuner,kerastuner}, and compare classification accuracy of the automatically optimized network to the non-optimized ones. Throughout all subsections, we also present network accuracy results using ``HLS-simulated'' versions of the CNNs, produced using the \textit{hls4ml} package \cite{Fahim:2021cic}.
One key feature of \textit{hls4ml} is a reduction in accuracy due to quantization of the network, which we avoid by employing quantization-aware training, following \cite{Coelho:2020zfu,Hawks:2021ruw}, as discussed in Sec.~\ref{subsec:QCNN_DS}. Finally, in Sec.~\ref{sec:eval}, we provide estimates of FPGA resource usage of the optimized networks (with and without quantization-aware training), using an \textit{hls4ml} synthesized design for a targeted FPGA hardware implementation. We demonstrate that the use of 2D CNNs for real-time data selection in the future DUNE is viable, and advantageous, given the envisioned front-end readout system design.

\section{Application Case: Real-time Data Selection for the Future DUNE FD LArTPC}
\label{sec:dune}

\begin{figure}
    \centering
    \includegraphics[width=12cm]{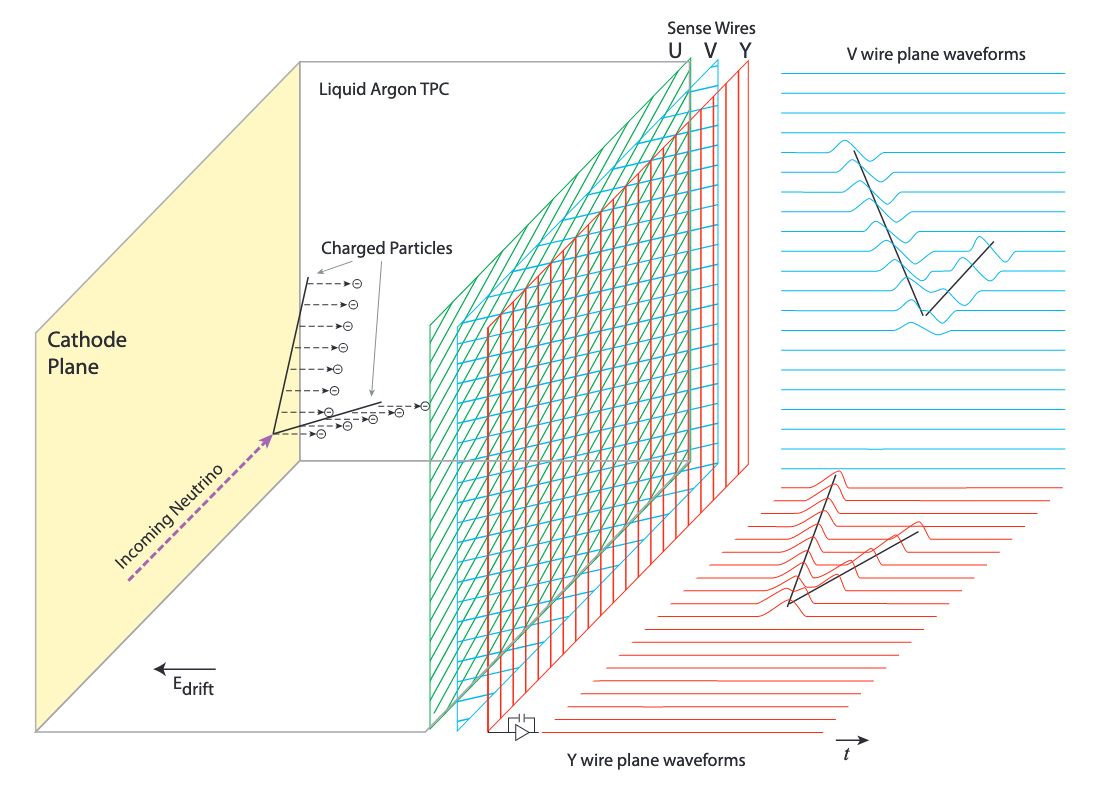}
    \caption{Operating principle of a LArTPC. The ionization electrons are drifted toward sensor arrays, e.g.~planes of ionization charge sensor wires. Each wire is connected to an analog amplifier/shaper, followed by an ADC, and its resulting digital waveform is read out continually. Waveforms of adjacent wires appended together form 2D images. Image credit: \cite{MicroBooNE:2016pwy}.}
    \label{fig:lartpc}
\end{figure}

Liquid Argon Time Projection Chambers (LArTPCs) are a state-of-the-art charged-particle detector technology with broad applications in the field of particle physics, astro-particle physics, nuclear physics, and beyond. This high-rate imaging detector technology has been adopted by multiple particle physics experiments, including the current MicroBooNE experiment \cite{MicroBooNE:2016pwy}, two additional detectors that are part of the upcoming Short-Baseline Neutrino (SBN) program \cite{MicroBooNE:2015bmn}, as well as the next-generation DUNE experiment \cite{DUNE:2020lwj,DUNE:2020mra,DUNE:2020txw,DUNE:2020ypp}, and it is also proposed for future-generation astro-particle physics experiments such as GRAMS \cite{Aramaki:2019bpi}. LArTPCs work by imaging ionization electrons produced along the paths of charged particles, as they travel through a large (multiple cubic meters) volume of liquid argon. Charged particle ionization trails drift uniformly toward sensor arrays with the use of a uniform electric field applied throughout the liquid argon volume, and are subsequently read out in digital format as part of 2D projected views of the 3D argon volume. This is illustrated in Fig.~\ref{fig:lartpc}. The densely packed sensor arrays sample the drifted ionization charge at a high rate, typically using a 12-bit, 2~MHz Analog to Digital Converted (ADC) system recording the amount of ionization charge per sensor per time-sample, thus imaging charge deposition across 2D projections of the argon volume with millimeter-scale resolution. Typically, digitized image frames of O(10)~megabytes each are streamed out of these detectors in real time and at a rate of up to hundreds of thousands of frames per second, amounting to raw data rates of multiple gigabytes to several terabytes per second. 


The future DUNE experiment presents a special case, with the most stringent data processing requirements among all currently running or planned LArTPC experiments. DUNE consists of a near and a far detector complex, which will be located at Fermi National Accelerator Laboratory (Fermilab) in Illinois and at the Sanford Underground Research Facility (SURF) in South Dakota, respectively. The far detector (FD) complex will be located 1 mile deep under ground, and will comprise the largest LArTPC ever to be constructed, with an anticipated raw data rate for its first of four LArTPC modules of 1.175~TB/s. This first detector module will be operated continually, and for at least ten years, starting as early as 2026, with subsequent modules coming online before the end of the current decade. The DUNE FD will therefore be constructed with a readout and data selection system that is required to receive and process an overall raw data rate of 4$\times$1.175~TB/s, achieve a factor of $10^4$ data reduction, and maintain $>99$\% efficiency to particle interactions of interest that are predicted to be as rare as once per century \cite{DUNE:2020txw}. 

The scientific goals of DUNE include, but are not limited to, observing neutrinos from rare (once per century) galactic supernova bursts (SNBs) \cite{DUNE:2020ypp,DUNE:2020zfm}, searching for rare baryon number violation processes such as argon-bound proton decay and argon-bound neutron-antineutron oscillation, and studying interactions of neutrinos that are produced in cosmic ray air showers in the earth's atmosphere \cite{DUNE:2020fgq,DUNE:2020ypp}. From the data acquisition (DAQ) and data selection (trigger) point of view, these rare physics searches and in particular the requirement to be $>99$\% efficient to a galactic SNBs with a less than once per month false positive SNB detection rate, cast particularly stringent technical requirements.

More specifically, in order to select rare physics events, which take place randomly and unpredictably, the DUNE DAQ and trigger system must scan \textit{all} detector data continuously and with zero dead time, and identify rare physics signatures of interest in a ``self-triggering'' mode---without relying on any external, early-warning signals prompting data readout in anticipation of a rare event. Furthermore, a self-triggering scheme reaching nearly perfect (100\%) efficiency for rare physics events is needed in order for DUNE to achieve its full physics reach. A particular challenge in triggering in this way is the need for temporarily buffering large amounts of data while processing it to make a data selection decision. In the case of DUNE, buffering constraints translate into a sub-second latency requirement for the trigger decision. Additionally, the trigger decision needs to achieve an overall $10^4$ data rate reduction, and with high signal selection efficiency, corresponding to an average of $>$60\% efficiency on individual supernova neutrino interactions, and $>$90\% efficiency to other rare interactions including atmospheric neutrino interactions and baryon number violating events.

The first DUNE FD module will image charged particle trajectories within 200 independent but contiguous liquid argon volume regions (``cells''). Charged particle trajectories within each cell will be read out by sensor wires arranged in three planes: one charge-collection wire planes, plus two charge-induction wire planes. Each plane read out corresponds to a particular 2D projected view of the 3D cell volume, and the combination of induction and collection plane information allows for 3D stereoscopic imaging and reconstruction of a given interaction within the 3D cell volume. For this work, we focus exclusively on charge-collection wire readout. Charge-collection wires give rise to signals which are unipolar in nature (as opposed to charge-induced signals, which are bipolar in nature, and therefore susceptible to cancellation effects). As such, charge-collection readout waveforms preserve sensitivity to charge deposition even for extended charge distributions. Since particle identification (and subsequent data selection decision making) relies on quantifying the amount of charge deposition per unit length of a charged particle track, charge-collection waveform information is anticipated to provide better particle identification performance. In total, the first FD module will consist of 384,000 wire sensors, each read out independently; this outnumbers current LArTPC neutrino experiments by more than a factor of 500 (e.g.~MicroBooNE makes use of 8,256 wire sensors). 

The 200 cells of the first DUNE FD module will be read out in parallel, by 75 ``upstream DAQ'' readout units. Each unit makes use of a Front-End LInk eXchange (FELIX) PCIe 3.0 card \cite{DUNE:2020txw,Borga:2018uqw} holding a Xilinx Virtex-7 UltraScale+ FPGA to read out digitized waveforms, and pre-process the data. In the nominal DUNE readout unit design, the FPGA processes continuous waveforms in order to perform noise filtering and hit-finding; hit-finding summaries are then sent for additional processing to a FELIX-host CPU system, in order to form trigger candidates (interaction candidates); the latter inform a subsequent module-wide trigger decision. An alternate potential solution, explored further through this work, is to apply more advanced data processing and triggering algorithms within the available FPGA resources on-board the FELIX card, such as CNNs, which can intelligently classify a collection of waveforms representing activity across the entire cell volume in real time, thus eliminating the need of subsequent CPU host (or GPU) processing, and potentially further minimizing power requirements. It is worth noting that, since most interactions of interest have a spatial extent which is smaller than the cell volume, a per-cell parallelization of triggering algorithms is appropriate, and it is therefore sufficient to focus trigger studies to a per-cell level, ignoring cell volume boundary effects.

\section{CNN Design and Optimization for Real-time LArTPC Data Selection}
\label{sec:CNNDS}

In recent years, ML algorithms such as CNNs have shown tremendous growth of their use in high energy physics analyses, including physics with LArTPCs \cite{Karagiorgi:2021ngt}. In particular, CNNs have been shown to achieve very high signal selection efficiencies especially when employed in offline physics analyses of LArTPC data. 
MicroBooNE is leading the development and application of ML techniques, including CNNs, for event reconstruction and physics analysis as an operating LArTPC \cite{MicroBooNE:2016dpb,MicroBooNE:2018kka,MicroBooNE:2020hho,MicroBooNE:2020yze}, and CNN-based analyses and ML-based reconstruction are actively being developed for SBN and for DUNE \cite{SBND:2020eho,DUNE:2020gpm}. 



In a previous study \cite{Jwa:2019zlh}, we have shown that sufficiently high efficiencies can be reached by processing raw collection plane data from any given DUNE FD cell, prior to removing any detector effects or applying data reconstruction. As such, we proposed a CNN-based triggering scheme using streaming raw 2D image frames, whereby the images are pre-processed, downsized, and run through CNN inference to select images (data) containing SNB neutrino interactions or other rare interactions of interest on a frame-by-frame basis. The data pre-processing and CNN-based selection method used demonstrated that target signal selection efficiency while reaching the needed $10^4$ background rejection could be achieved, given sufficient parallelization in GPUs. As the DUNE FD DAQ and trigger design is subject to stringent power limitations and limited accessibility in the underground detector cavern, a particularly attractive option is to fully implement this pre-processing and CNN-based inference on FPGAs, in particular ones that will be part of the DUNE upstream DAQ readout unit design. We examine the viability of this option in this work. 

Specifically, we explore the accuracy of relatively small CNNs in classifying streaming DUNE FD LArTPC cell data, and proceed to employ network optimization in an effort to reduce its computational resource footprint while preserving network accuracy. The following subsections describe the CNN input image preparation (Sec.~\ref{subsec:prep}), CNN performance without (Sec.~\ref{subsec:CNN_DS}) and with (Sec.~\ref{subsec:auto}) network optimization, and with quantization-aware training (Sec.~\ref{subsec:QCNN_DS}). 

\subsection{Input Image Pre-processing}
\label{subsec:prep}
\begin{figure}[h]
    \centering
    \includegraphics[width=15cm]{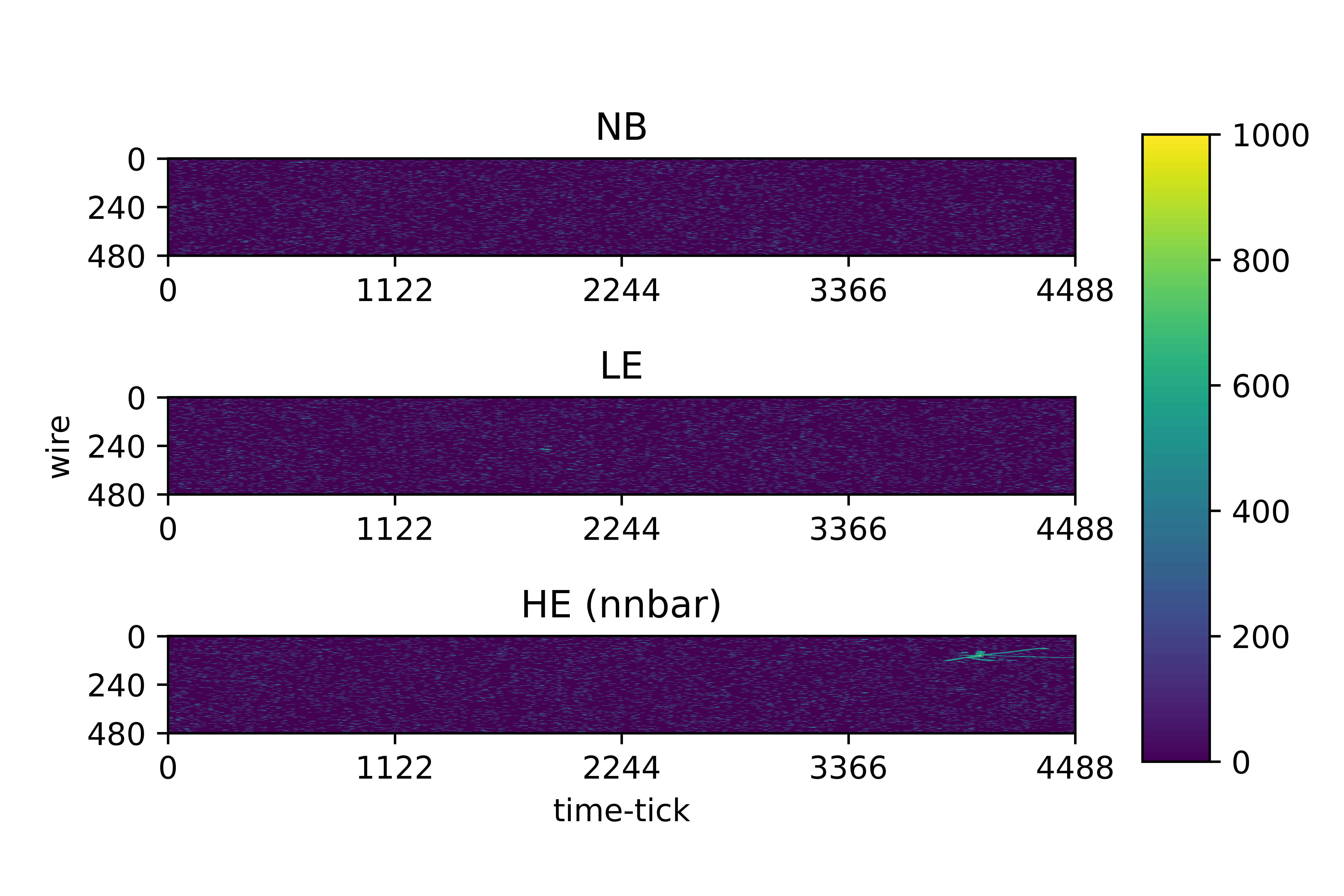}
    \caption{Examples of 2D images formed from one full drift (2.25ms) of 480 collection plane wires in one DUNE FD cell. Top: Image containing electronics noise and radiological background only (NB). Middle: Image containing one low-energy supernova neutrino interaction (LE) superimposed with electronics noise and radiological background. Bottom: Image containing one high-energy interaction (HE), specifically from neutron-antineutron oscillation (nnbar), superimposed with electronics noise and radiological background. These images are pre-processed prior to CNN processing.}
    \label{fig:fullscale_input}
\end{figure}
Because of the parallelism in the DUNE FD DAQ and trigger design, we only consider a single cell's worth of data at a time, and focus exclusively on raw collection plane waveforms. Following \cite{Jwa:2019zlh}, collection plane waveforms for a single cell in the DUNE FD are simulated in the LArSoft framework \cite{larsoft,Church:2013hea}, using the default configuration of the \emph{dunetpc} software version 7.13.00, and using an enhanced electronics noise level configuration, to be conservative. 
Besides electronics noise, the simulation includes radiological impurity background interactions that are intrinsic to the liquid argon volume. The radiological background interactions (predominantly from $^{39}$Ar decay) are expected to occur at a rate of $10^{7}$~Hz per FD module, and they are considered as likely backgrounds particularly to supernova neutrino interactions. Signal waveforms from interactions of interest, including low-energy supernova neutrino interactions or other high-energy interactions (proton decay, neutron-antineutron oscillation, atmospheric neutrino interactions, cosmogenic background interactions), are overlaid on top of intrinsic radiological background and electronics noise waveforms. 

Given the physical dimension of a cell along the ionization charge drift direction, and the known ionization charge drift velocity, 2.25~ms worth of continuous data from the entire collection plane represents a 2D image exposure of the full cell volume. As such, we define a 2D image in terms of 480 collection plane wire channels spanning the length of the cell volume, times the 2.25~ms drift direction sampled at 2~MHz (4488 samples) spanning the width of the cell volume. This corresponds to a 2.1 megapixel image, with 12-bit ADC resolution governing the range of pixel values, dictating the amount of ionization charge collected by each wire, and indicating the energy deposit within the 3D volume along the given 2D projection.

For network training purposes, the 2.1 megapixel input images are labeled as containing either electronics noise and radiological background only (NB), or low-energy supernova neutrino interactions (LE) superimposed with electronics noise and radiological background, or high-energy interactions (HE) superimposed with electronics noise and radiological background, according to the simulation truth. Figure~\ref{fig:fullscale_input} shows example input 2D images before pre-processing steps. We note the sparsity of these images, mostly containing uniformly distributed low-energy activity from noise and radiological backgrounds. While it is possible to train a CNN with 2.1 megapixel images, it is not memory-efficient, and it may furthermore not be an efficient way to propel a CNN to learn the different features between the three event classes (NB, LE, and HE). Following \cite{Jwa:2019zlh}, we adopt pre-processing steps that include de-noising (zero-supression), cropping around the region-of-interest (ROI), and resizing the ROI through down- or up-sampling. The de-noising step uses a configurable threshold for the pixel ADC value and zero-suppresses pixel values below this threshold; a threshold of 520 ADC (absolute scale) was used in these studies, where $\sim500$ ADC represents the baseline. ROI cropping was performed by finding a contiguous rectangular region containing pixels with values over 560 ADC. The most extreme image coordinates (smallest and largest channel number, as well as smallest and largest time tick) with pixel values greater than the lower threshold of 560 ADC were used to determine the ROI boundaries. Once an ROI was found, the ROI region was resized (through up-sampling or down-sampling) to occupy exactly $64\times64$ pixels. 

\begin{figure}[h]
    \centering
    \includegraphics[width=4.5cm]{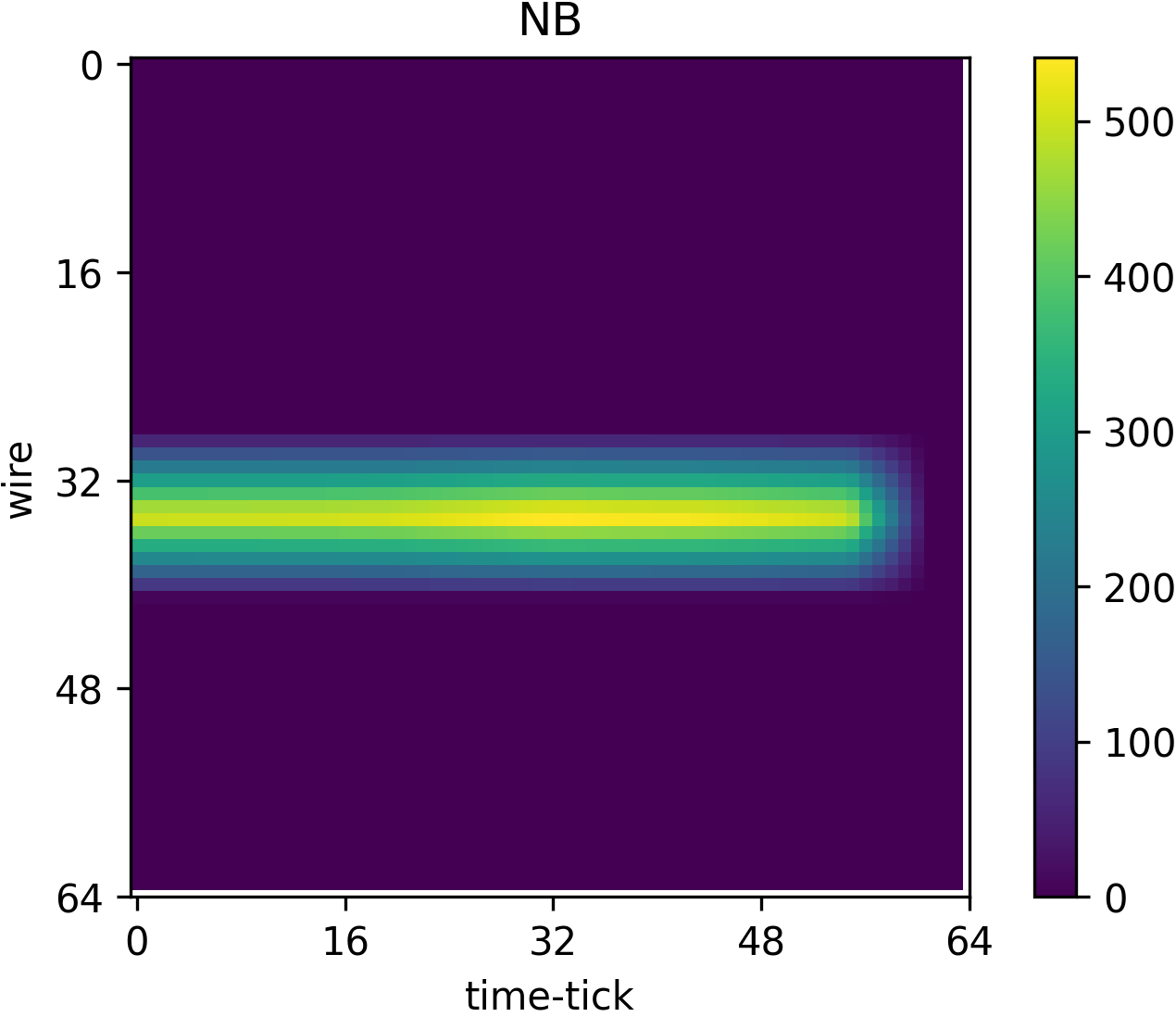}
    \includegraphics[width=4.5cm]{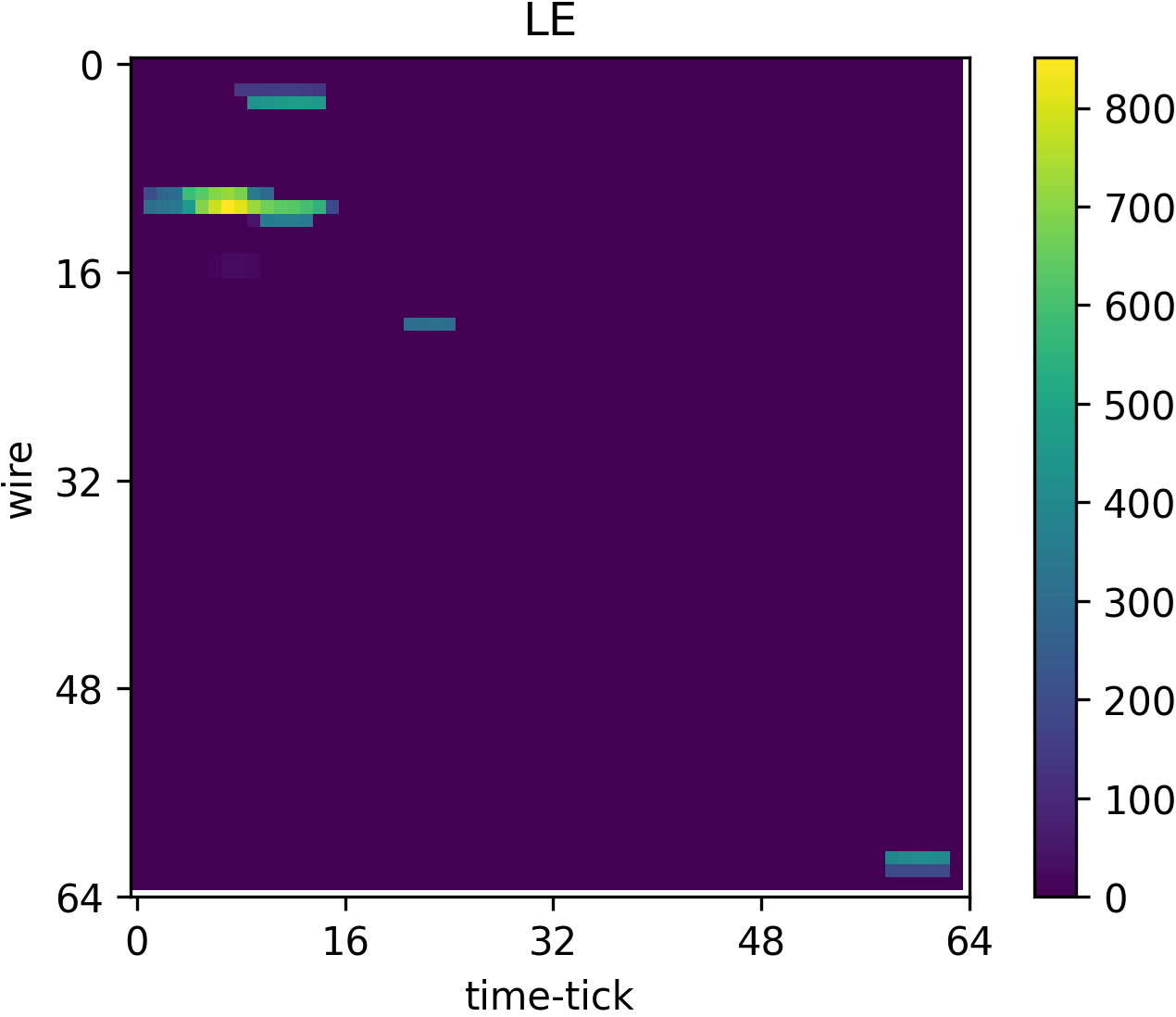}
    \includegraphics[width=4.5cm]{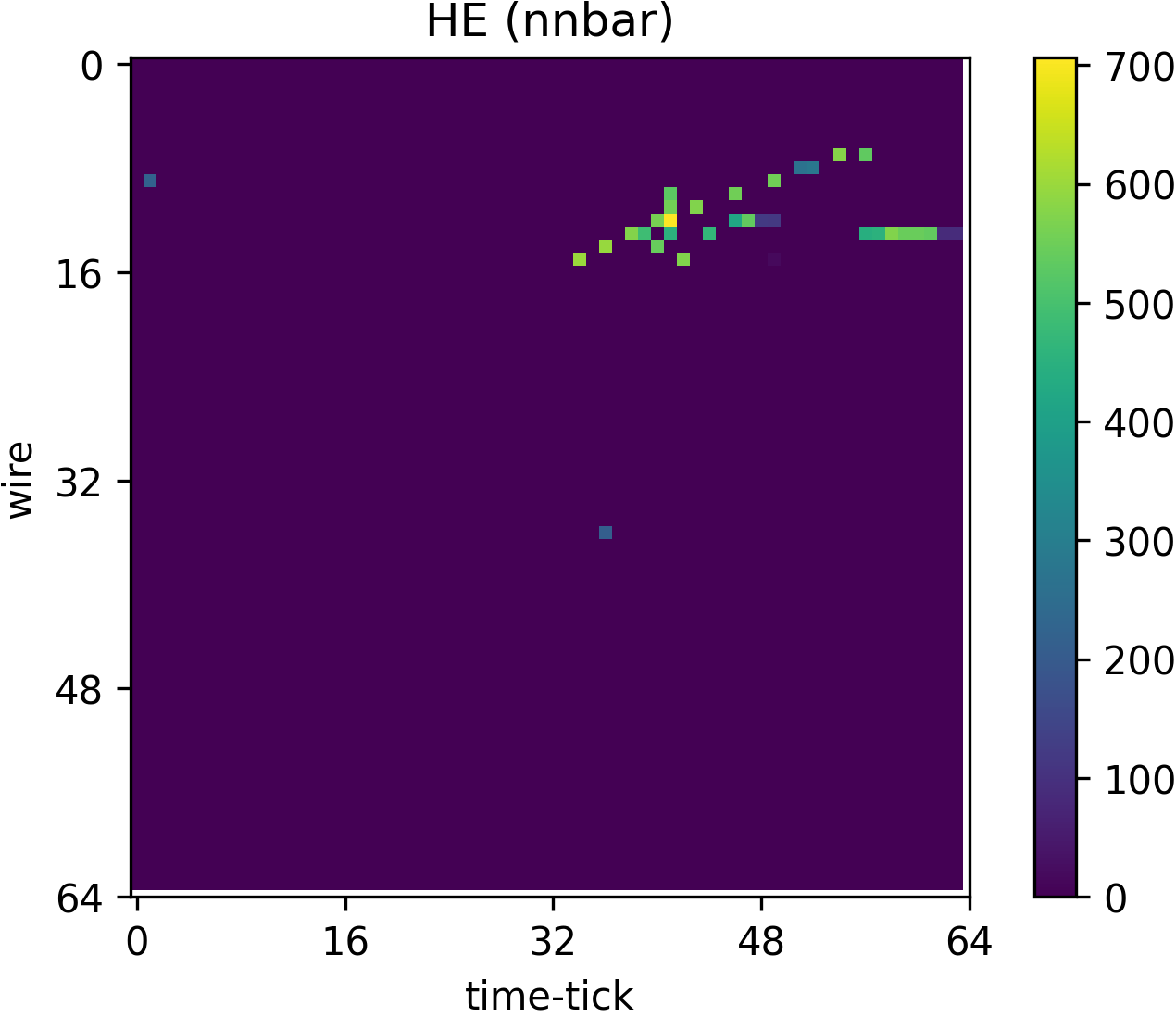}
    \caption{Example ROIs formed after pre-processing. Left: Image containing electronics noise and radiological background only (NB). Middle: Image containing one low-energy supernova neutrino interaction (LE) superimposed with electronics noise and radiological background. Right: Image containing one high-energy interaction (HE), specifically from neutron-antineutron oscillation (nnbar), superimposed with electronics noise and radiological background. These images are input to a CNN for subsequent processing (data selection).}
    \label{fig:ROIs}
\end{figure}

Resized ROIs were generated for each of the three categories indicated in Tab.~\ref{tab:ROIs}, with comparable statistics, and used for network training and testing for all studies presented in the subsequent sections. A total of 45,624 ROIs were used in the study. A 75\%:25\% split was used for training:testing sets.

\begin{table}[ht]
\caption{Number of ROIs, according to truth label, used for training and testing of CNNs. A total of 45,624 ROIs were used in the study. A 75\%:25\% split was used for training:testing sets.}
\begin{center}
\label{tab:ROIs}
\begin{tabular}{lrrr}
\hline\hline
&Label: NB&Label: LE&Label: HE\\
\hline
Training set size&12,023&12,050&10,137\\
Testing set size&4,027&3,970&3,417\\
\hline\hline
\end{tabular}
\end{center}
\end{table}

\begin{figure}
    \centering
    \includegraphics[width=10cm]{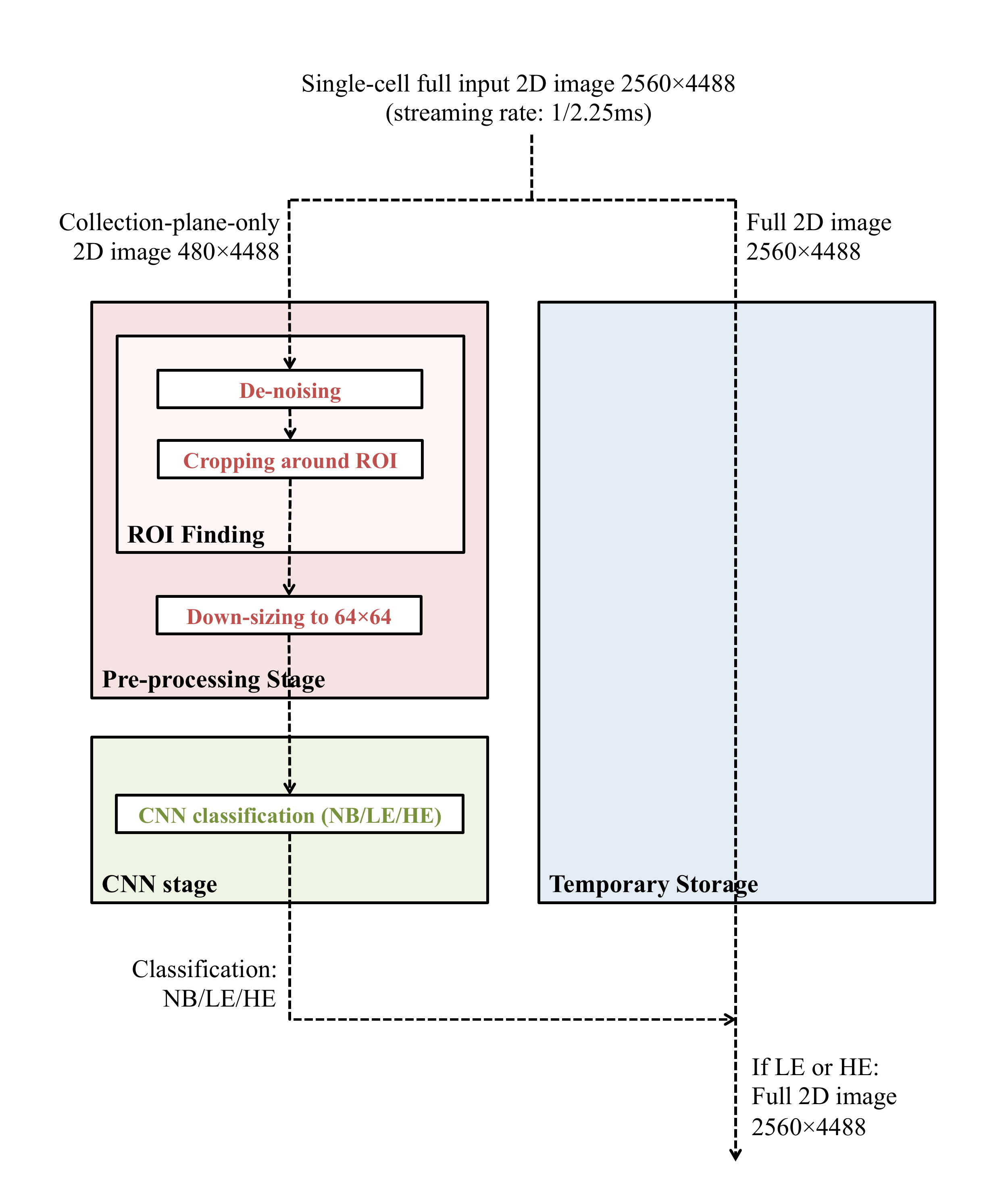}
    \caption{The data processing and data selection scheme under study for potential implementation in the upstream DAQ readout units of the future DUNE FD. The streaming 2D input images contain, $>99.9$\% of the time, NB data. This overall scheme should select true HE and LE images with $>90$\% accuracy, and true NB images with $>99.99$\% accuracy, in order to meet the DUNE FD physics requirements. Additionally, the pre-processing and CNN inference algorithms should meet the computational resources of the DUNE FD upstream DAQ readout units, and the algorithm execution latency should meet the data throughput requirements of the experiment.}
    \label{fig:dataflow}
\end{figure}

The overall data processing and data selection scheme proposed and examined in this study is summarized in Fig.~\ref{fig:dataflow}. 

\subsection{Performance of CNN-based Data Selection}
\label{subsec:CNN_DS}
Targeting FPGA implementation, we designed and tested custom CNN architectures with one or two convolutional layers: \textbf{CNN01}, \textbf{CNN02}, and a downsized version of the latter, \textbf{CNN02-DS}. These networks have far simpler architectures than some of the more popular CNN architectures commonly used in image classification tasks (e.g.~VGG or ResNet network architectures), by design, as they are targeted for implementation in computational-resource-constrained systems.

\begin{figure}[h]
    \centering
    \includegraphics[width=12cm]{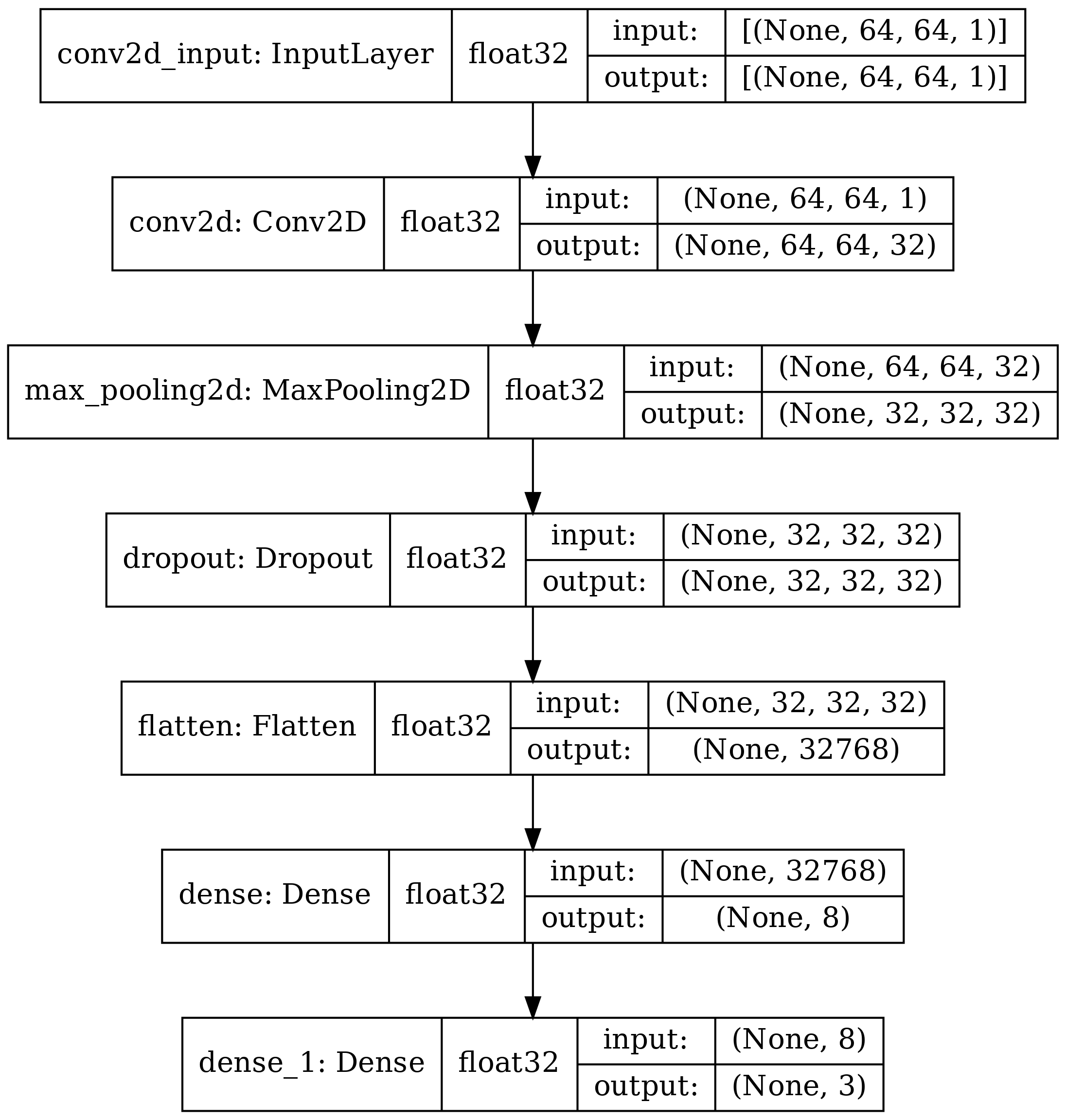}
    \caption{Network architecture of \textbf{CNN01}.}
    \label{fig:CNN01architecture}
\end{figure}

The network architecture for \textbf{CNN01} is shown in Fig.~\ref{fig:CNN01architecture}. \textbf{CNN01} has one convolutional layer, with convolutional width kernel dimension (3,3,32), and one max-pooling layer. One fully connected layer follows at the end. In contrast, \textbf{CNN02} has two convolutional layers, and one max-pooling layer after each convolution. Also, here, one fully connected layer follows at the end. Finally, \textbf{CNN02-DS} is a downsized version of \textbf{CNN02}, where the convolution depth is significantly reduced. All three custom network architectures are summarized in Tab.~\ref{tab:CNNs}.

\begin{table}[ht]
\caption{Summary of explored CNN architectures.}
\begin{center}
\begin{tabular}{lccc}
\hline\hline
&\bf{CNN01}&\bf{CNN02}&\bf{CNN02-DS}\\
\hline
Number of convolution layers&1&2&2\\
Convolution kernel dimension (first conv.)&3$\times$3$\times$32&3$\times$3$\times$32&3$\times$3$\times$4\\
Convolution kernel dimension (second conv.)&N/A&3$\times$3$\times$64&3$\times$3$\times$8\\
Number of max-pooling layers&1&2&2\\
Max-pooling dimension (first max-pool)&2$\times$2&2$\times$2&4$\times$4\\
Max-pooling dimension (second max-pool)&N/A&2$\times$2&4$\times$4\\
Number of trainable parameters& 262,499&149,923&1,395\\
\hline\hline
\end{tabular}
\label{tab:CNNs}
\end{center}
\end{table}

Table \ref{tab:keras_classifications} shows the classification performance of the three networks, for a GPU or CPU implementation using Keras \cite{keras}. 
The performance of these three networks is comparable. For all three networks, the false positive identification rates (which affect data reduction capability) are comparable, and the (correct) classification accuracy is over 99\% for NB labeled ROIs, over 93\% for LE labeled ROIs, and over 90\% for HE labeled ROIs. Despite the difference in architecture (one vs.~two convolution layers) and number of trainable parameters, no clear impact on classification performance is observed.

\begin{table}[ht]
\caption{Classification accuracy comparison for \textbf{CNN01}, \textbf{CNN02}, and \textbf{CNN02-DS} on GPU or CPU.}
\begin{center}
\label{tab:keras_classifications}
\begin{tabular}{lrrr}
\hline\hline
\bf{CNN01}&NB&LE&HE\\
\hline
true\_NB&99.45\%&0.55\%&0\%\\
true\_LE&3.83\%&94.23\%&1.94\%\\
true\_HE&3.37\%&6.14\%&90.49\%\\
\hline\hline
\bf{CNN02}&NB&LE&HE\\
\hline
true\_NB& 99.5\%& 0.5\%& 0\%\\
true\_LE& 3.98\%& 93.25\%& 2.77\%\\
true\_HE& 3.25\%& 6.58\%& 90.17\%\\
\hline\hline
\bf{CNN02-DS}&NB&LE&HE\\
\hline
true\_NB&99.48\%&0.52\%&0\%\\
true\_LE&3.70\%&94.38\%&1.92\%\\
true\_HE&3.01\%&6.50\%&90.49\%\\
\hline\hline
\end{tabular}
\end{center}
\end{table}

While accuracy results meet signal efficiency requirements\footnote{In this study, accuracy is defined identically to signal efficiency, i.e.~as a true positive classification rate given a set of true labels.}, the high false positive rate (in particular for true NB ROIs to be mis-classified as LE events at a rate of 0.5\%) suggests a steady-state data reduction factor for a frame-by-frame data selection implementation that is a factor of 50 lower than the required reduction factor of 10$^4$. This is because the overwhelming majority (>99.9\%) of the streaming ROIs in DUNE are expected to be truly NB ROIs, and therefore a 0.5\% mis-classification rate would result in approximately one in 200 ROIs being (falsely) selected, as opposed to the targeted one in 10,000. Additional data reduction, however, can be provided by an ROI pre-selection stage, as motivated in \cite{Jwa:2019zlh}; specifically, approximately only one in 50 2D true NB images are expected to be non-empty after ROI finding (see Fig.~\ref{fig:dataflow}) and therefore 98\% of the ROIs can be discarded prior to CNN processing.\footnote{Note that the CNN studies presented in this paper are performed exclusively on non-empty ROIs. For images containing LE and HE events, all images survive after ROI-finding; i.e., ROI-finding does not cause any additional reduction in efficiency, and the ROI classification accuracy represents the signal efficiency. For images containing only NB, only one in approximately 50 images survive after ROI-finding.} This suggests that an overall factor of 10$^4$ is achievable. 

In this work, the ML models were trained and tested on GPUs with single-precision floating-point arithmetic (standard IEEE 754), and then post-training quantization (PTQ) was performed with the aim of running ML inference on FPGA. It is worth noting that FPGAs support integer, floating-point, and fixed-point arithmetic. An FPGA implementation may require orders of magnitude higher resources, besides higher latency and power costs, when compared with a finely-tuned fixed-point implementation of the same algorithm~\cite{finnerty2017}. Predictably, PTQ impacts ML classification performance, although the profiling tools in \textit{hls4ml} help the designer decide the appropriate model precision~\cite{hls4ml-profiling}.
The resulting accuracy values for PTQ networks targeted for FPGA (with fixed-point precision) are shown in Tab.~\ref{tab:hls_sim}, and contrasted to those with floating-point precision in Tab.~\ref{tab:Post-training-Quantization}. We adopted quantization-aware training (QAT) to address this accuracy drop, as discussed in Sec.~\ref{subsec:QCNN_DS}.

\begin{table}[ht]
\begin{center}
\caption{Classification accuracy comparison for \textbf{CNN01}, \textbf{CNN02}, and \textbf{CNN02-DS}, using post-training quantization (PTQ).}
\label{tab:hls_sim}
\begin{tabular}{lrrr}
\hline\hline
\bf{CNN01}&NB&LE&HE\\
\hline
true\_NB&98.14\%&1.84\%&0.02\%\\
true\_LE&6.57\%&89.10\%&4.33\%\\
true\_HE&19.37\%&37.72\%&42.90\%\\
\hline\hline
\bf{CNN02}&NB&LE&HE\\
\hline
true\_NB& 98.06\%& 0.25\%& 1.91\%\\
true\_LE& 22.75\%& 10.6\%& 66.65\%\\
true\_HE& 21.54\%& 3.72\%& 74.74\%\\
\hline\hline
\bf{CNN02-DS}&NB&LE&HE\\
\hline
true\_NB&99.53\%&0.47\%&0\%\\
true\_LE&4.94\%&93.12\%&1.94\%\\
true\_HE&21.22\%&40.12\%&38.66\%\\
\hline\hline
\end{tabular}
\end{center}
\end{table}

\begin{table}[ht]
\caption{Combined classification accuracy for true NB, LE, and HE ROIs for floating-point vs.~PTQ fixed-point implementations of the trained networks. }
\begin{center}
\label{tab:Post-training-Quantization}
\begin{tabular}{lrrr}
\hline\hline
 &\bf{CNN01}&\bf{CNN02}&\bf{CNN02-DS}\\
\hline
Floating-point accuracy&94.95\%&94.53\%&95.01\%\\
Fixed-point accuracy (PTQ) &78.46\%&60.66\%&79.08\%\\
\hline\hline
\end{tabular}
\end{center}
\end{table}


\subsection{Automatized CNN Hyperparameter Optimization using KerasTuner}
\label{subsec:auto}

In the initial network performance comparison presented in Sec.~\ref{subsec:CNN_DS}, the classification performance does not appear to be highly sensitive to the network architecture and number of trainable parameters; further optimization of networks with respect to a large phase-space of hyperparameters can be performed methodically and in an automated way using open-source tools such as KerasTuner, as described below.

\begin{table}[ht]
\caption{Scanning range and granularity of the hyperparameters explored during automated network optimization using KerasTuner.}
\begin{center}
\label{tab:hyperparameters}
\begin{tabular}{lcc}
\hline\hline
\textbf{Hyperparameter}&\textbf{Range}&\textbf{Default Value}\\
\hline
First convolution depth (conv1)& $[4, 8, 16]$ &4\\
Second convolution depth (conv2)& $[8,16, 32]$&8\\
Dense layer size (fc)& $[8, 12, 16, 20, 24]$&12\\
Learning rate (lr), logarithmic sampling& $[2\times10^{-4}, 2\times10^{-2}]$ & $2\times10^{-3}$\\
\hline\hline
\end{tabular}
\end{center}
\end{table}

The choice of network hyperparameters such as the the dimensions of hidden layers, and learning parameters, changes the number of trainable variables. Thus, the quality of training can be modulated by tuning the hyperparameters.
Hyperparameter tuning can be done manually, but it is a cumbersome procedure to tweak hyperparameters and compare the classification performance in a controlled way.
KerasTuner is an open-source hyperparameter optimization framework that solves the pain points of hyperparameter search, and was used for hyperparameter optimization for the baseline network architecture \textbf{CNN02-DS}. The scanning range and granularity of the hyperparameters explored is shown in Tab.~\ref{tab:hyperparameters}.
A total of twenty combinations were randomly sampled from the hyperparameter scanning region, with the results from the five best-performing combinations and the default configuration  shown in Tab.~\ref{tab:optimization}. 
The optimized network \textbf{CNN02-DS-OP} with the highest classification accuracy found at 95.221\%, corresponds to a network with a first convolution depth of 8, second convolution depth of 16, dense layer size of 12 , and learning rate of $2.9 \times 10^{-3}$. 

\begin{table}[h!]
\caption{Classification accuracy for the five top-performing and default (\textbf{CNN02-DS}) hyperparameter configurations. Note that the default accuracy obtained during hyperparameter optimization slightly differs from that in Tab.~\ref{tab:Post-training-Quantization}, due to differences in (random) initialization of the network weights before training, and randomness during the training.}
\begin{center}
\label{tab:optimization}
\begin{tabular}{lccccc}
\hline\hline
&\textbf{conv1}&\textbf{conv2}&\textbf{fc}&\textbf{lr}& \textbf{accuracy}\\
\hline
First-best (\textbf{CNN02-DS-OP})&8&16&12&2.9 $\times 10^{-3}$& 95.221\%\\
Second-best&16&32&12&4.9 $\times 10^{-4}$& 95.206\%\\
Third-best&4&16&20&6.0 $\times 10^{-4}$& 95.206\%\\
Fourth-best&16&8&16&7.0$ \times 10^{-4}$& 95.191\%\\
Fifth-best&16&8&12&1.9$ \times 10^{-3}$& 95.191\%\\
default&4&8&12&2$ \times 10^{-3}$&95.089\%\\
\hline\hline
\end{tabular}
\end{center}
\end{table}

\subsection{Network Quantization in CNN-based Data Selection}
\label{subsec:QCNN_DS}

The cost reduction and performance improvement of fixed-point arithmetic with HLS is highly encouraged when designing ML algorithms for FPGA deployment. Typically, when a trained network within an ML framework (e.g.~Keras) on CPU or GPU is translated to HLS, the floating-point precision is reduced to the fixed-point precision of a given configuration. As a consequence, generally, network quantization resulting from fixed-point precision effectively reduces the precision of the calculations for weights, bias, and/or inputs, resulting in lower inference accuracy performance than what would otherwise be possible with floating-point precision. This is evident in Tab.~\ref{tab:Post-training-Quantization}. 

In principle, one cannot achieve the flexibility and accuracy of a floating-point precision with any fixed-point representation. However, if accuracy can be maintained with an optimized choice of fixed-point precision,
one can benefit from the inherent advantage of reduced computing resource utilization. This (maintaining of accuracy) can be achieved with quantization-aware network training \cite{Coelho:2020zfu,Hawks:2021ruw}.

\begin{figure}[h!!]
    \centering
    \includegraphics[width=10cm]{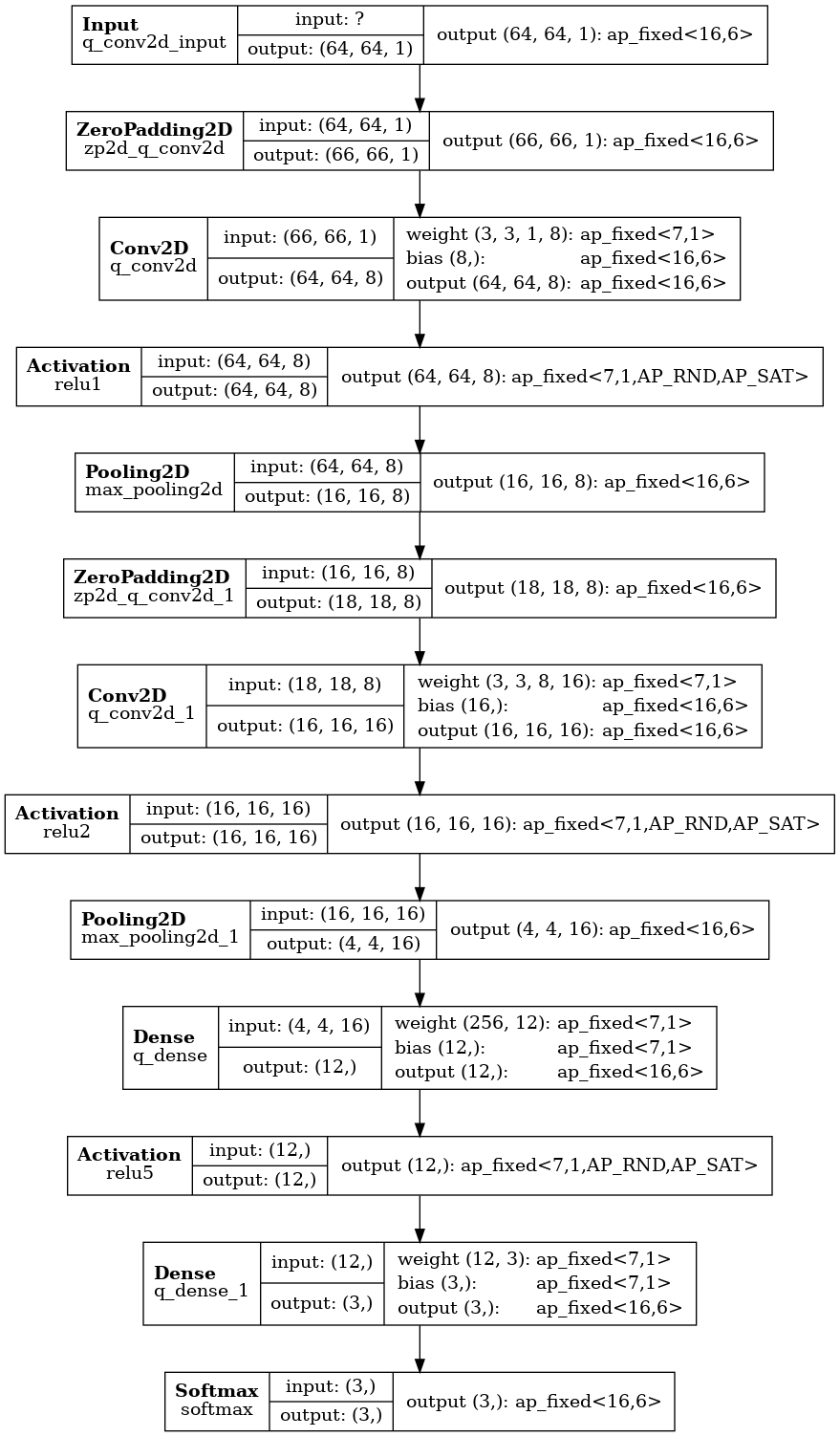}
    \caption{Precision configuration of layers in \textbf{Q-CNN02-DS-OP}. The precision configuration of the reference \textbf{CNN02-DS-OP} can be found in Fig.~\ref{fig:CNN_config}.}
    \label{fig:QCNN_config}
\end{figure}
\begin{figure}[ht]
    \centering
    \includegraphics[width=10cm]{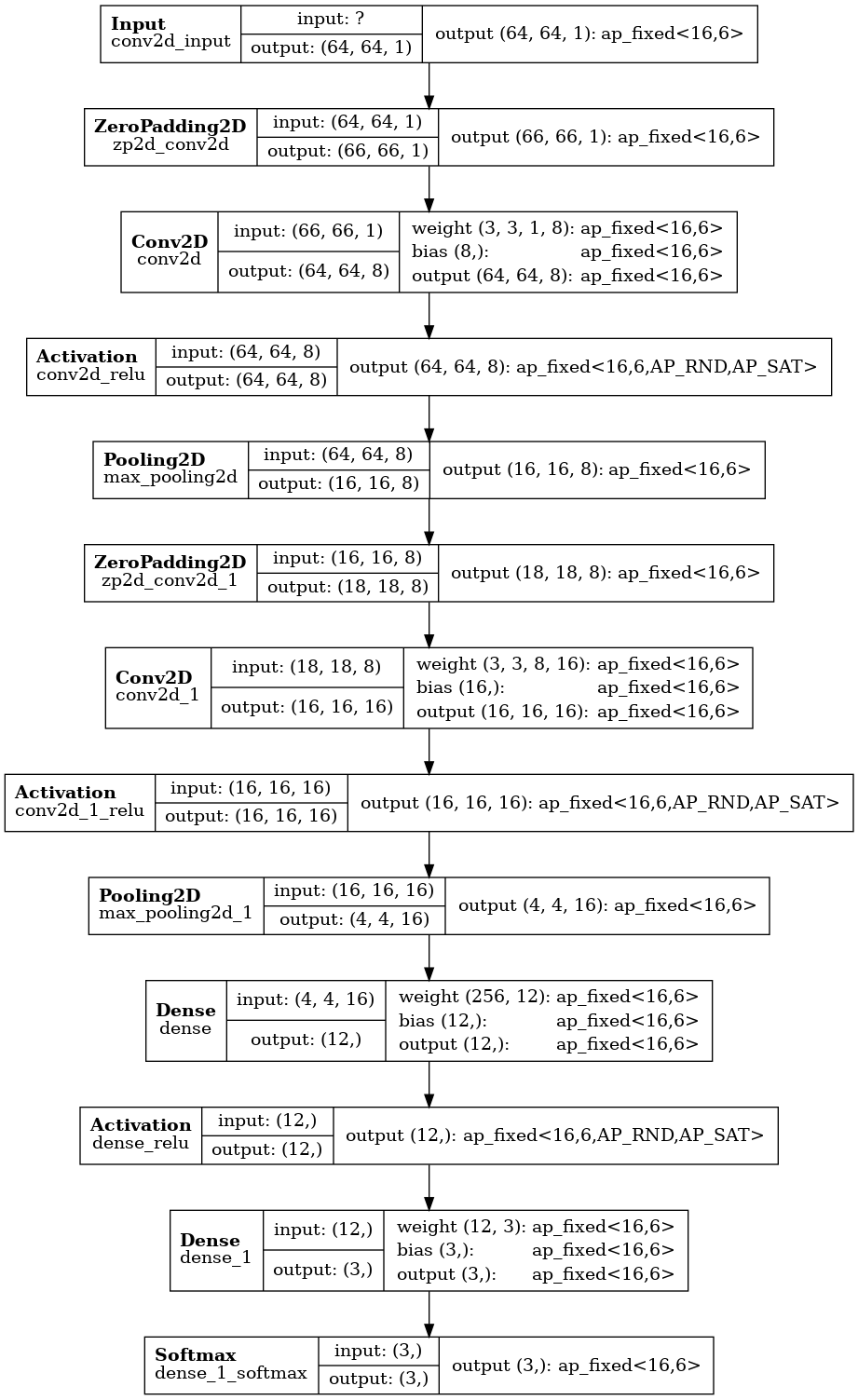}
    \caption{Precision configuration of layers in the reference \textbf{CNN02-DS-OP}.}
    \label{fig:CNN_config}
\end{figure}

Quantization-aware training (QAT), achieved by committing calculations in ML algorithms with already-reduced fixed-point representation as part of network training, can prevent reduction in inference accuracy.
The QKeras package \cite{qkeras} supports quantization-aware training by quantizing any given network using Qlayers, such as QActivation, QDense, Qconv2D, etc.
The quantized network derived from a given network architecture can be constructed by replacing the layers in the initial network to Qlayers. We refer to the quantized version of \textbf{CNN02-DS-OP} obtained with QKeras as 
\textbf{Q-CNN02-DS-OP}. The precision configuration of \textbf{Q-CNN02-DS-OP} is shown in Fig.~\ref{fig:QCNN_config}. The precision configuration of the reference \textbf{CNN02-DS-OP} is shown is shown in Fig.~\ref{fig:CNN_config}.

The classification results obtained using the reference \textbf{CNN02-DS-OP} network with and without PTQ are shown in Tab.~\ref{tab:PTQ}; the corresponding results obtained with the quantization-aware trained (QAT) \textbf{Q-CNN02-DS-OP} are shown in Tab.~\ref{tab:QAT}. 

\begin{table}[ht]
\caption{Optimized performance for \textbf{CNN02-DS-OP}, without quantization-aware network training.}
\begin{center}
\label{tab:PTQ}
\begin{tabular}{lccc}
\hline\hline
\textbf{Floating-point}&NB&LE&HE\\
\hline
true\_NB&99.50\%&0.50\%&0\%\\
true\_LE&3.53\%&95.09\%&1.39\%\\
true\_HE&2.93\%&6.12\%&90.96\%\\
\hline
\multicolumn{1}{l}{total accuracy} & \multicolumn{3}{c}{95.41\%}\\
\hline\hline
\textbf{Fixed-point (PTQ)}&NB&LE&HE\\
\hline
true NB&  99.83\%&  0.17\%& 0\%\\
true LE&  6.32\%&  88.87\%&  4.81\%\\
true HE&  26.89\%&  52.15\%& 20.95\%\\
\hline
\multicolumn{1}{l}{total accuracy} & \multicolumn{3}{c}{72.40\%}\\
\hline\hline
\end{tabular}
\end{center}

\caption{Optimized performance for \textbf{Q-CNN02-DS-OP}, with quantization-aware network training.}
\begin{center}
\label{tab:QAT}
\begin{tabular}{lccc}
\hline\hline
\textbf{Floating-point (QAT)}&NB&LE&HE\\
\hline
true\_NB& 99.60\%& 0.40\%& 0\%\\
true\_LE& 3.78\%& 94.03\%& 2.19\%\\
true\_HE& 3.19\%& 5.44\%& 91.37\%\\
\hline
\multicolumn{1}{l}{total accuracy} & \multicolumn{3}{c}{ 95.20\%}\\
\hline\hline
\textbf{Fixed-point (QAT)}&NB&LE&HE\\
\hline
true\_NB& 99.68\%& 0.32\%& 0\%\\
true\_LE& 3.90\%& 94.69\%& 1.41\%\\
true\_HE& 3.25\%& 6.35\%& 90.40\%\\
\hline
\multicolumn{1}{l}{total accuracy} & \multicolumn{3}{c}{ 95.16\%}\\
\hline\hline
\end{tabular}
\end{center}
\end{table}

For the network trained without QAT, \textbf{CNN02-DS-OP}, the overall classification accuracy for the entire testing sample (superset of three truth labels) drops significantly with PTQ, from 95.41\% to 72.40\%.
For the network trained with QAT, \textbf{Q-CNN02-DS-OP}, however, the overall classification accuracy is maintained for what would be an equivalent FPGA implementation (with PTQ), at 95.20\% and 95.16\%. This demonstrates that a relatively small CNN, applied on a frame-by-frame basis, and trained with quantization that is consistent with FPGA fixed-point precision, can achieve the accuracy (signal efficiency and target data reduction factor) required for the DUNE FD. 

\section{Estimation of FPGA Resource Usage}
\label{sec:eval}

In this section, we estimate FPGA resource usage and examine whether a Xilinx Virtex-7 UltraScale+ FPGA can comfortably accommodate a pre-trained CNN that meets the accuracy as well as resource and latency specifications of the DUNE FD DAQ and trigger system.

The estimated hardware usage for the quantized inference block of each of the optimized CNNs (\textbf{Q-CNN02-DS-OP} and \textbf{CNN02-DS-OP}) from Vivado HLS is shown in Tab.~\ref{tab:resource}. The hardware usage of the discussed inference shows that the target FPGA, a high-end device, is well fit for implementing either the \textbf{Q-CNN02-DS-OP} or the \textbf{CNN02-DS-OP} network. As expected, the \textbf{Q-CNN02-DS-OP} network uses significantly lower FPGA resources. It is worth noting that, in addition to using more resources, \textbf{CNN02-DS-OP} (PTQ) has a lower accuracy than \textbf{Q-CNN02-DS-OP} (QAT), at 72.4\% vs.~95.16\%, illustrating the advantages of QAT.

Assuming a clock cycle of 5.00~ns, we find that the design is expected to meet timing requirements, with an inference latency of 
4680 clock-cycles, corresponding to 23.4~$\mu$s. This is well below the exposure time corresponding to a single input image of 2.25~ms; thus, assuming sufficient parallelization (i.e.~at least two input 2D images processed in parallel by each FELIX unit), frame-by-frame real-time data selection based on collection plane-only image analysis with CNNs is a viable solution for the DUNE FD. Note that this does not consider additional resource utilization or latency associated with image pre-processing (ROI finding and down-sizing).

\begin{table}[h]
\caption{Estimated resource utilization from Vivado HLS for CNN inference on a Xilinx UltraScale+ (XCKU115) FPGA. }
\begin{center}
\label{tab:resource}
\begin{tabular}{lllll}
\hline\hline
&\textbf{Block RAM}&\textbf{DSP Units}&\textbf{Flip Flops} &\textbf{Look-up Tables}\\
\hline
Available&4320&5520&1326720&663360\\
\hline
\textbf{CNN02-DS-OP (PQT)}&331 (7\%)&4309 (78\%)&226982 (17\%)&163460 (24\%)\\
\textbf{Q-CNN02-DS-OP (QAT)}&187 (4\%)&2106 (38\%)&142128 (10\%)&138715 (20\%)\\
\hline\hline
\end{tabular}
\end{center}
\end{table}

We note that, in the current stage, the FPGA design has been synthesized, but it has not been implemented yet into the hardware; this is the focus of continuing development efforts. 

\section{Summary}

In recent years, ML algorithms such as CNNs have shown tremendous growth of their use in high energy physics, including physics analysis with LArTPCs \cite{Karagiorgi:2021ngt}. In particular, CNNs have been shown to achieve very high signal selection efficiencies especially when employed in offline physics analyses of LArTPC data. 
MicroBooNE is leading the development and application of ML techniques, including CNNs, for event reconstruction and physics analysis as an operating LArTPC \cite{MicroBooNE:2016dpb,MicroBooNE:2018kka,MicroBooNE:2020hho,MicroBooNE:2020yze}, and CNN-based analyses and ML-based reconstruction are actively being developed for SBN and for DUNE \cite{SBND:2020eho,DUNE:2020gpm}. 



Motivated by a previous study \cite{Jwa:2019zlh}, showing that CNN-based data selection for LArTPC detectors can yield excellent accuracy even when applied solely at raw collection plane data, we have propose a 2D CNN-based, real-time, frame-by-frame data selection (trigger) scheme that is a viable solution for the DUNE FD. Leveraging the extensive parallelization available within the DUNE FD upstream DAQ readout design, in this proposed scheme, 2D image frames streamed at a total rate of 1.25~TB/s are pre-processed and run through CNN inference to classify and select interactions of interest on a frame-by-frame basis. The proposed pre-processing and CNN-based selection method yield target signal selection efficiencies that meet the DUNE FD physics requirements, while also providing the needed $10^4$ factor of overall data rate reduction.

The FPGA resource utilization for the CNN inference has been optimized with automatized network optimization and with quantization-aware training so as to avoid accuracy loss due to a fixed-point precision implementation in FPGA. The resulting optimized and quantized CNN (\textbf{Q-CNN02-DS-OP}) has been shown to fit within available DUNE FD upstream DAQ readout FPGA resources, and to be executable with sufficiently low latency such that the need for significant buffering resources in the DUNE FD upstream DAQ system can also be relaxed. We note, however, that the pre-processing resource requirements and latency have not been explicitly evaluated, and this will be the subject of future work, as they need to be considered in tandem with the proposed CNN algorithm. 




The findings further motivate future LArTPC readout designs that preserve the physical mapping of readout channels to a contiguous interaction volume as much as possible, in order to minimize pre-processing needs, and preserve spatial correlations that exist within 2D projected views of the interaction volume. Additionally, they motivate the consideration of other image analysis algorithms in the designs of DAQ and trigger systems of future LArTPCs. 

\acknowledgments
This work is based upon work supported by the National Science Foundation under Grant No.~NSF-1914965.


\bibliographystyle{unsrt}

\begin{thebibliography}{}


\bibitem{Karagiorgi:2021ngt}
G.~Karagiorgi, G.~Kasieczka, S.~Kravitz, B.~Nachman and D.~Shih,
``Machine Learning in the Search for New Fundamental Physics,''
[arXiv:2112.03769 [hep-ph]].


\bibitem{Radovic:2018dip}
A.~Radovic, M.~Williams, D.~Rousseau, M.~Kagan, D.~Bonacorsi, A.~Himmel, A.~Aurisano, K.~Terao and T.~Wongjirad,
``Machine learning at the energy and intensity frontiers of particle physics,''
Nature \textbf{560}, no.7716, 41-48 (2018)
doi:10.1038/s41586-018-0361-2


\bibitem{Jwa:2019zlh}
Y.~J.~Jwa, G.~D.~Guglielmo, L.~P.~Carloni and G.~Karagiorgi,
``Accelerating Deep Neural Networks for Real-time Data Selection for High-resolution Imaging Particle Detectors,''
doi:10.1109/NYSDS.2019.8909784
[arXiv:2201.04740 [physics.ins-det]].

\bibitem{Elabd:2021lgo}
A.~Elabd, V.~Razavimaleki, S.~Y.~Huang, J.~Duarte, M.~Atkinson, G.~DeZoort, P.~Elmer, J.~X.~Hu, S.~C.~Hsu and B.~C.~Lai, \textit{et al.}
``Graph Neural Networks for Charged Particle Tracking on FPGAs,''
[arXiv:2112.02048 [physics.ins-det]].


\bibitem{Diotalevi:2021vlw}
T.~Diotalevi \textit{et al.} [CMS],
``Deep Learning fast inference on FPGA for CMS Muon Level-1 Trigger studies,''
PoS \textbf{ISGC2021}, 005 (2021)
doi:10.22323/1.378.0005


\bibitem{Aad:2021tru}
G.~Aad, A.~S.~Berthold, T.~Calvet, N.~Chiedde, E.~M.~Fortin, N.~Fritzsche, R.~Hentges, L.~A.~O.~Laatu, E.~Monnier and A.~Straessner, \textit{et al.}
``Artificial Neural Networks on FPGAs for Real-Time Energy Reconstruction of the ATLAS LAr Calorimeters,''
Comput. Softw. Big Sci. \textbf{5}, no.1, 19 (2021)
doi:10.1007/s41781-021-00066-y

\bibitem{Govorkova:2021utb}
E.~Govorkova, E.~Puljak, T.~Aarrestad, T.~James, V.~Loncar, M.~Pierini, A.~A.~Pol, N.~Ghielmetti, M.~Graczyk and S.~Summers, \textit{et al.}
``Autoencoders on FPGAs for real-time, unsupervised new physics detection at 40 MHz at the Large Hadron Collider,''
[arXiv:2108.03986 [physics.ins-det]].

\bibitem{Mikuni:2021nwn}
V.~Mikuni, B.~Nachman and D.~Shih,
``Online-compatible Unsupervised Non-resonant Anomaly Detection,''
[arXiv:2111.06417 [cs.LG]].


\bibitem{Deiana:2021niw}
A.~M.~Deiana, N.~Tran, J.~Agar, M.~Blott, G.~Di Guglielmo, J.~Duarte, P.~Harris, S.~Hauck, M.~Liu and M.~S.~Neubauer, \textit{et al.}
``Applications and Techniques for Fast Machine Learning in Science,''
[arXiv:2110.13041 [cs.LG]].

\bibitem{Fahim:2021cic}
F.~Fahim, B.~Hawks, C.~Herwig, J.~Hirschauer, S.~Jindariani, N.~Tran, L.~P.~Carloni, G.~Di Guglielmo, P.~Harris and J.~Krupa, \textit{et al.}
``hls4ml: An Open-Source Codesign Workflow to Empower Scientific Low-Power Machine Learning Devices,''
[arXiv:2103.05579 [cs.LG]].

\bibitem{Loncar:2020hqp}
V.~Loncar, J.~Ngadiuba, J.~Duarte, P.~Harris, D.~Hoang, V.~Loncar, K.~Pedro, M.~Pierini, D.~Rankin and S.~Sagear, \textit{et al.}
``Compressing deep neural networks on FPGAs to binary and ternary precision with HLS4ML,''
Mach. Learn. Sci. Tech. \textbf{2}, 015001 (2021)
doi:10.1088/2632-2153/aba042
[arXiv:2003.06308 [cs.LG]].

\bibitem{Aarrestad:2021zos}
T.~Aarrestad, V.~Loncar, N.~Ghielmetti, M.~Pierini, S.~Summers, J.~Ngadiuba, C.~Petersson, H.~Linander, Y.~Iiyama and G.~Di Guglielmo, \textit{et al.}
``Fast convolutional neural networks on FPGAs with hls4ml,''
Mach. Learn. Sci. Tech. \textbf{2}, no.4, 045015 (2021)
doi:10.1088/2632-2153/ac0ea1
[arXiv:2101.05108 [cs.LG]].

\bibitem{DUNE:2020lwj}
B.~Abi \textit{et al.} [DUNE],
``Deep Underground Neutrino Experiment (DUNE), Far Detector Technical Design Report, Volume I Introduction to DUNE,''
JINST \textbf{15}, no.08, T08008 (2020)
doi:10.1088/1748-0221/15/08/T08008
[arXiv:2002.02967 [physics.ins-det]].

\bibitem{DUNE:2020ypp}
B.~Abi \textit{et al.} [DUNE],
``Deep Underground Neutrino Experiment (DUNE), Far Detector Technical Design Report, Volume II: DUNE Physics,''
[arXiv:2002.03005 [hep-ex]].

\bibitem{DUNE:2020mra}
B.~Abi \textit{et al.} [DUNE],
``Deep Underground Neutrino Experiment (DUNE), Far Detector Technical Design Report, Volume III: DUNE Far Detector Technical Coordination,''
JINST \textbf{15}, no.08, T08009 (2020)
doi:10.1088/1748-0221/15/08/T08009
[arXiv:2002.03008 [physics.ins-det]].

\bibitem{DUNE:2020txw}
B.~Abi \textit{et al.} [DUNE],
``Deep Underground Neutrino Experiment (DUNE), Far Detector Technical Design Report, Volume IV: Far Detector Single-phase Technology,''
JINST \textbf{15}, no.08, T08010 (2020)
doi:10.1088/1748-0221/15/08/T08010
[arXiv:2002.03010 [physics.ins-det]].


\bibitem{omalley2019kerastuner}
    O'Malley, Tom and Bursztein, Elie and Long, James and Chollet, Fran\c{c}ois and Jin, Haifeng and Invernizzi, Luca and others, ``KerasTuner'', \url{https://github.com/keras-team/keras-tuner}, 2019.
    
    
\bibitem{kerastuner}
``KerasTuner'', \url{https://keras.io/keras_tuner/}. Accessed: December 20, 2021.

    
\bibitem{Coelho:2020zfu}
C.~N.~Coelho, A.~Kuusela, S.~Li, H.~Zhuang, T.~Aarrestad, V.~Loncar, J.~Ngadiuba, M.~Pierini, A.~A.~Pol and S.~Summers,
``Automatic heterogeneous quantization of deep neural networks for low-latency inference on the edge for particle detectors,''
doi:10.1038/s42256-021-00356-5
[arXiv:2006.10159 [physics.ins-det]].

\bibitem{Hawks:2021ruw}
B.~Hawks, J.~Duarte, N.~J.~Fraser, A.~Pappalardo, N.~Tran and Y.~Umuroglu,
``Ps and Qs: Quantization-aware pruning for efficient low latency neural network inference,''
doi:10.3389/frai.2021.676564
[arXiv:2102.11289 [cs.LG]].


\bibitem{MicroBooNE:2016pwy}
R.~Acciarri \textit{et al.} [MicroBooNE],
``Design and Construction of the MicroBooNE Detector,''
JINST \textbf{12}, no.02, P02017 (2017)
doi:10.1088/1748-0221/12/02/P02017
[arXiv:1612.05824 [physics.ins-det]].

\bibitem{MicroBooNE:2015bmn}
M.~Antonello \textit{et al.} [MicroBooNE, LAr1-ND and ICARUS-WA104],
``A Proposal for a Three Detector Short-Baseline Neutrino Oscillation Program in the Fermilab Booster Neutrino Beam,''
[arXiv:1503.01520 [physics.ins-det]].


\bibitem{Aramaki:2019bpi}
T.~Aramaki, P.~Hansson Adrian, G.~Karagiorgi and H.~Odaka,
``Dual MeV Gamma-Ray and Dark Matter Observatory - GRAMS Project,''
Astropart. Phys. \textbf{114}, 107-114 (2020)
doi:10.1016/j.astropartphys.2019.07.002
[arXiv:1901.03430 [astro-ph.HE]].

\bibitem{DUNE:2020zfm}
B.~Abi \textit{et al.} [DUNE],
Eur. Phys. J. C \textbf{81}, no.5, 423 (2021)
doi:10.1140/epjc/s10052-021-09166-w
[arXiv:2008.06647 [hep-ex]].

\bibitem{DUNE:2020fgq}
B.~Abi \textit{et al.} [DUNE],
``Prospects for beyond the Standard Model physics searches at the Deep Underground Neutrino Experiment,''
Eur. Phys. J. C \textbf{81}, no.4, 322 (2021)
doi:10.1140/epjc/s10052-021-09007-w
[arXiv:2008.12769 [hep-ex]].


\bibitem{Borga:2018uqw}
A.~Borga, E.~Church, F.~Filthaut, E.~Gamberini, P.~de Jong, G.~Lehmann Miotto, F.~Schreuder, J.~Schumacher, R.~Sipos and M.~Vermeulen, \textit{et al.}
``FELIX based readout of the Single-Phase ProtoDUNE detector,''
IEEE Trans. Nucl. Sci. \textbf{66}, no.7, 993-997 (2019)
doi:10.1109/TNS.2019.2904660
[arXiv:1806.09194 [physics.ins-det]].


\bibitem{MicroBooNE:2016dpb}
R.~Acciarri \textit{et al.} [MicroBooNE],
``Convolutional Neural Networks Applied to Neutrino Events in a Liquid Argon Time Projection Chamber,''
JINST \textbf{12}, no.03, P03011 (2017)
doi:10.1088/1748-0221/12/03/P03011
[arXiv:1611.05531 [physics.ins-det]].

\bibitem{MicroBooNE:2018kka}
C.~Adams \textit{et al.} [MicroBooNE],
``Deep neural network for pixel-level electromagnetic particle identification in the MicroBooNE liquid argon time projection chamber,''
Phys. Rev. D \textbf{99}, no.9, 092001 (2019)
doi:10.1103/PhysRevD.99.092001
[arXiv:1808.07269 [hep-ex]].


\bibitem{MicroBooNE:2020hho}
P.~Abratenko \textit{et al.} [MicroBooNE],
``Convolutional neural network for multiple particle identification in the MicroBooNE liquid argon time projection chamber,''
Phys. Rev. D \textbf{103}, no.9, 092003 (2021)
doi:10.1103/PhysRevD.103.092003
[arXiv:2010.08653 [hep-ex]].


\bibitem{MicroBooNE:2020yze}
P.~Abratenko \textit{et al.} [MicroBooNE],
``Semantic segmentation with a sparse convolutional neural network for event reconstruction in MicroBooNE,''
Phys. Rev. D \textbf{103}, no.5, 052012 (2021)
doi:10.1103/PhysRevD.103.052012
[arXiv:2012.08513 [physics.ins-det]].

\bibitem{SBND:2020eho}
R.~Acciarri \textit{et al.} [SBND],
``Cosmic Background Removal with Deep Neural Networks in SBND,''
[arXiv:2012.01301 [physics.data-an]].


\bibitem{DUNE:2020gpm}
B.~Abi \textit{et al.} [DUNE],
``Neutrino interaction classification with a convolutional neural network in the DUNE far detector,''
Phys. Rev. D \textbf{102}, no.9, 092003 (2020)
doi:10.1103/PhysRevD.102.092003
[arXiv:2006.15052 [physics.ins-det]].
LaTeX (US)


\bibitem{Church:2013hea}
E.~D.~Church,
``LArSoft: A Software Package for Liquid Argon Time Projection Drift Chambers,''
[arXiv:1311.6774 [physics.ins-det]].


\bibitem{larsoft}
``LArSoft'', \url{https://larsoft.org/}. Accessed: December 20, 2021.


\bibitem{keras}
    ``Keras'', \url{https://github.com/keras-team/keras/}. Accessed: December 20, 2021.


\bibitem{finnerty2017}
Finnerty, Ambrose and Ratigner, and Herv{\'e}, 
``Reduce power and cost by converting from floating point to fixed point'', Xilinx WP491, 2017.

\bibitem{hls4ml-profiling}
    ``Profiling'', \url{https://fastmachinelearning.org/hls4ml/api/profiling.html}. Accessed: January 2, 2022.

\bibitem{qkeras}
    ``QKeras'', \url{https://github.com/google/qkeras}. Accessed: December 20, 2021.








\end{thebibliography}

\clearpage

\end{document}